\documentclass[journal=jacsat,manuscript=article,layout=twocolumn]{achemso}

\usepackage{chemformula} 
\usepackage[T1]{fontenc} 

\title{Comment on ``Loss-Free Excitonic Quantum Battery''}
\author{\'Alvaro Tejero}
\affiliation{Universidad de Granada, Departamento de Electromagnetismo y Física de la Materia and Instituto Carlos I de Física Teórica y
Computacional, Granada 18071, Spain.}

\author{Juzar Thingna}
\affiliation{Center for Theoretical Physics of Complex Systems, Institute for Basic Science (IBS), Daejeon 34126, Republic of Korea.}
\alsoaffiliation{Basic Science Program, University of Science and Technology (UST), Daejeon 34113, Republic of Korea.}
\email{jythingna@ibs.re.kr}

\author{Daniel Manzano}
\affiliation{Universidad de Granada, Departamento de Electromagnetismo y Física de la Materia and Instituto Carlos I de Física Teórica y
Computacional, Granada 18071, Spain.}
\email{manzano@onsager.ugr.es}

\begin{document}
Quantum batteries have primarily been modeled as an ensemble of isolated systems that store energy and from which work can be extracted by applying unitary transformations~\cite{campaioli}. Only recently, investigations have begun in the direction of \emph{dissipative} quantum batteries \cite{liu,barra,quach} wherein the charge of the battery is protected against a dissipative environment. 

In Ref.~[2], a novel model for a quantum battery is proposed based on a degenerated quantum system with a topological symmetry. The working substance consists of a para-Benzene ring which is initially prepared in a dark-state, $\rho_{dark} = |DS\rangle\langle DS|$ with $|DS\rangle = \frac{1}{2}\left(|5\rangle + |6\rangle - |2\rangle - |3\rangle \right)$, considered as the ``charged'' state of the battery. The geometric symmetries of the para-Benzene are translated to open system symmetries~\cite{Thingna16} since the environment selectively acts on sites $1$ and $4$ independently. The authors claim that after breaking the system's symmetry, by the addition of an external probe, the energy in the system is redistributed, and it is transferred to  site $4$, where it can be harnessed by connecting a sink. The system dynamics are calculated in the single-excitation picture by a mixed quantum-classical method known as the deterministic evolution of coordinates with initial decoupled equations (DECIDE) \cite{DECIDE}. The authors present only the system dynamics but do not discuss the energetics (e.g., ergotropy), which is essential to understand a quantum battery's proper functioning. Moreover, the simulations do not show how exactly the energy can be extracted using a sink in this complex set up. We would like to point out that the capacity of a system of providing excitations is not enough to be considered a battery. There are examples of passive states (e.g. states whose ergotropy is zero and therefore they cannot provide any useful work)~\cite{campaioli} that present population imbalance in site representation. 

We do not dispute their approach or their result on the single-excitation populations. We argue that observing only the single-excitation populations is insufficient to know the energetics. Specifically, since an infinite probe is attached to break the symmetries, it not only redistributes the energies as the authors suggest but exchanges energy with the system. Moreover, the accumulation of a single-excitation population at site $4$ does not ensure that this site has more energy (from a many-body perspective) that can be extracted to do useful work. Thus, despite their single-excitation populations being correct, the para-Benzene's energetics do not behave as a battery.

\begin{figure*}
\includegraphics[scale=1]{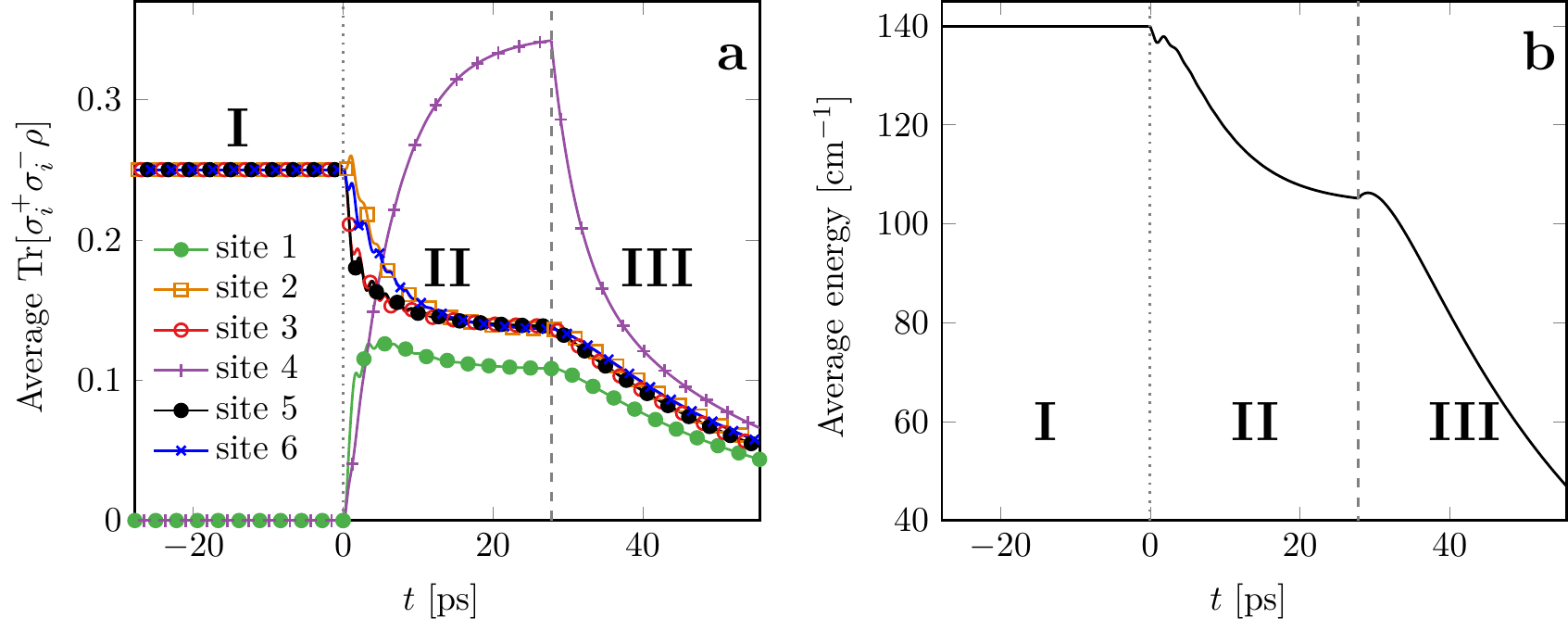}
\caption{Many-body Redfield results for time-dependent populations and average energy. The initial conditions for region I is the dark state $\rho_{dark}$ and for region III is the steady state of region II. The parameters for the system are: $\epsilon=200\mathrm{cm}^{-1}$, $\epsilon_1 = 250\mathrm{cm}^{-1}$, $\epsilon_4=0\mathrm{cm}^{-1}$, $t=-60\mathrm{cm}^{-1}$, $\lambda_b = 35\mathrm{cm}^{-1}$, $\lambda_p=10\mathrm{cm}^{-1}$, $\Gamma=0.1$, $\omega_c=\omega_p=106\mathrm{cm}^{-1}$ (Lorentz-Drude spectral density cut-off), $T=T_p=300$K (chosen same as Fig.~1 in Ref.~[2]).}
\label{fig1}
\end{figure*}
In order to capture the energetics we have modelled the same system in the many-body fully quantum picture ($\hbar=1$) using the Markovian Redfield equation~\cite{Redfield},
\begin{eqnarray}
\label{eq:1}
\frac{d\rho}{dt} &=& -i[H,\rho] + \sum_{k=1,4,\atop p}\mathcal{R}_k[\rho] + \mathcal{L}_{sink}[\rho],
\end{eqnarray}
where the dissipator $\mathcal{R}_k$ is given by
\begin{eqnarray}
\mathcal{R}_k[\rho]= \int_0^{\infty} d\tau [S_k,\rho S_k (-\tau)] C_{k}(\tau) +\mathrm{H.c.}
\end{eqnarray}
Above  $S_k(\tau) = e^{iH\tau}S_k e^{-iH\tau}$ is the freely-evolving system operator that connects to the bath, $C_{k}(\tau)= \mathrm{Tr}[B_{k}B_{k}(\tau)e^{-\beta_k H_B^k}/Z_B]$ is the correlation function of the bath with $Z_B$ being the partition function of the bath at inverse temperature $\beta_k$ and $B_{k}(\tau)=e^{iH_B^{k}\tau}B_k e^{-iH_B^{k}\tau}$. Here, $H_B^{k}$ is the Hamiltonian of the $k$th bath, $H_B^{p}$ is the probe Hamiltonian, $S_k = \sigma^+_k\sigma^-_k$ for $k=1,4$ and $S_{p}=\sqrt{\chi}\left(\sigma^+_2\sigma^-_2 + \sigma^+_3\sigma^-_3\right)$. These operators are the generalization of the ones used in Ref.~[2] to the multiple-excitations general case. We want also to point out that the baths and the probe preserve the total number of excitations of the system as they commute with the total number operator defined as $N=\sum_{i=1}^6\sigma_i^+ \sigma_i^-$. Thus, in the many-body representation even with a probe we obtain seven steady states each corresponding to the number of particles in the system ranging from 0-6. In this comment we will choose $\beta_1=\beta_2=[k_B T]^{-1}=\beta_{p}=[k_B T_p]^{-1}$ similar to Ref.~[2]. We also attach a sink $\mathcal{L}_{sink}[\rho] = \Gamma\left(\sigma^-_4\rho\sigma^+_4 - \frac{1}{2}\{\sigma^+_4\sigma^-_4,\rho\}\right)$ in order to extract energy in the discharging phase which breaks the particle-number conserving symmetry giving a single unique steady state. The many-body system Hamiltonian reads
\begin{equation}
H = \sum_{i=1,\cdots,6}\epsilon_i \sigma_i^+\sigma_i^{-} + t\sum_{\langle i,j\rangle} \left( \sigma_i^{+}\sigma_j^{-} +\mathrm{H.c.} \right),
\end{equation}
with $\sigma^{\pm}_i$ being spin-$\frac{1}{2}$ Pauli matrices. As in the single-excitation picture, the system presents a symmetry defined by the unitary operator $\Pi = \exp [i\pi\left(\sigma_1^+\sigma_1^- + \sigma_4^+\sigma_4^- + \left(\sigma_2^+\sigma_6^- + \sigma_3^+\sigma_5^- + \mathrm{H.c.}\right)\right)]$. 

We have calculated the populations for each site, defined as $\text{Tr} \left[ \sigma_i^+\sigma_i^-\rho \right]$, as well as the average energy of the system. The populations found in Fig.~2 of Ref.~[2] are very similar to the ones we obtain in Fig.~\ref{fig1}a. It is important to note here that instead of this definition of populations, we could have chosen the projection of the many-body reduced density matrix $\rho$ in the single-excitation subspace, i.e., $\langle i|\rho| i \rangle$. However, these again show the same behaviour (not shown). The imbalance between sites $1$ and $4$ is mainly due to the counter-rotating terms that are not ignored in the Redfield dynamics \cite{note1}. Besides the populations we also calculate the average energy of the battery $\text{Tr}\left[H\rho\right]$ in Fig.~\ref{fig1}b. Note that the average energy of the battery is not given by the sum of site populations weighted by the on-site energy, i.e.,$\sum_{i}\epsilon_i\text{Tr} \left[ \sigma_i^+\sigma_i^-\rho \right]$. This is because the energy is also stored between the sites due to the hopping $t$ and site-coherences.

The system evolution is divided into three different regimes that correspond to the regions in Fig.~\ref{fig1}. Region I is in without a probe or sink ($\chi = 0$ and $\Gamma=0$), Region II is with only a probe  ($\chi=1$ and $\Gamma=0$), and Region III is with both probe and a sink ($\chi=1$ and $\Gamma=0.1$). Once the probe is attached (region II), the system's energy decreases, and it is not merely rearranged to the various sites. In terms of battery operation, energy is extracted from the system (working substance) during the discharging phase $\chi=1$ instead of being rearranged to be ready for extraction. The energy reduction is without a load (sink) being attached, meaning this would result in a ``leaky'' battery. Once the sink is attached to site $4$, the exit site's population decreases as claimed in Ref.~[2]. Since the bath and the probe are kept at the same temperature the resultant $t\rightarrow \infty$ state is passive which would make it impossible to extract work even if unitary transformations (instead of attaching a sink) are performed in the discharging phase~\cite{campaioli}.

From the above analysis, we conclude that when the symmetry is broken (discharging phase), there is not just an energy redistribution in the system, but the probe exchanges energy with the system. For the number conserving probe we find that the attachment of the probe makes it a \emph{leaky} battery. The sink takes an infinite time to completely drain the battery since energy is provided by the probe and baths. We have tried other number non-conserving probes and found the probe can even give energy to the system. In no case was the energy of the system conserved except for pure-dephasing probe which would not change the populations of the system. Our analysis based on the exact many-body fully quantum dynamics and energetics raises serious doubts on whether the para-Benzene system proposed in Ref.~[2] is a practical loss-free quantum battery. 

Finally, we want also to point out that the symmetry operator in Eq.~(9) of Ref.~[2], earlier stated in Ref.~[5], is wrong as it is proportional to the identity operator. The correct one is
\begin{equation}
\hat{\Pi} = \left| 1 \right> \left< 1 \right| + \left| 4 \right> \left< 4 \right| + \left(\left| 2 \right> \left< 6 \right| + \left| 3 \right> \left< 5 \right| + \mathrm{H.c.}\right)
\end{equation}


\begin{thebibliography}{9}

\bibitem{campaioli}
F. Campaioli, F. A. Pollock, and S. Vinjanampathy. {\it Quantum Batteries}, in {\it Thermodynamics in the Quantum Regime}. Springer {\bf 2018}. 

\bibitem{liu} 
J. Liu, D. Segal, and G. Hanna. {\it Loss-Free Excitonic Quantum Battery}. J. Phys. Chem C {\bf 2019}, 123, 18303. 

\bibitem{barra}
F. Barra. {\it Dissipative Charging of a Quantum Battery}. Phys. Rev. Lett. {\bf 2019}, 122, 210601. 

\bibitem{quach}
J. Q. Quach and W. J. Munro. {\it  Using Dark States to Charge and Stabilize Open Quantum Batteries}. Phys. Rev. App. {\bf 2020}, 14, 024092.

\bibitem{Thingna16}
J. Thingna, D. Manzano, and J. Cao. {\it Dynamical Signatures of Molecular Symmetries in Nonequilibrium Quantum Transport}. Sci. Rep. {\bf 2016}, 6, 28027.

\bibitem{DECIDE}
J. Liu and G. Hanna, {\it Deterministic Propagation of Mixed Quantum-Classical Liouville Dynamics.} J. Phys. Chem. Lett. {\bf 2018}, 9, 3928. 

%

\bibitem{Redfield}
A. G. Redfield. {\it On the Theory of Relaxation Processes}. IBM J. Res. Dev. {\bf 1957}, 1, 19.

\bibitem{note1}
We also performed calculations using a Lindblad equation that neglects counter-rotating terms and found that the populations of sites $1$ and $4$ are always the same.

\end{thebibliography}
\end{document}